\documentclass[aps,onecolumn,showpacs,preprintnumbers,amsmath,amssymb,nofootinbib]{revtex4}
\usepackage{amssymb}
\usepackage{epsfig}
\usepackage{graphicx}

\usepackage{graphicx}
\usepackage{dcolumn}
\usepackage{bm}

\begin{document}

\title{Radiative E1 decays of X(3872)}

\author{Tian-Hong Wang}
\email{thwang.hit@gmail.com}
\author{Guo-Li Wang}
\email{gl_wang@hit.edu.cn}

\affiliation{Department of Physics, Harbin Institute of
Technology, Harbin 150001, China}

\baselineskip=20pt

\begin{abstract}
Radiative E1 decay widths of $\rm X(3872)$ are calculated through
the relativistic Salpeter method, with the assumption that $\rm
X(3872)$ is the $\chi_{c1}$(2P) state, which is the radial excited
state of $\chi_{c1}$(1P). We first calculated the E1 decay width of
$\chi_{c1}$(1P). The result is in agreement with experimental data
excellently. Then we calculated the width of $\rm X(3872)$ with the
assignment $\chi_{c1}$(2P). Results are: ${\Gamma}({\rm
X(3872)}\rightarrow \gamma \sl J/\psi)=33.0$ keV, ${\Gamma}({\rm
X(3872)}\rightarrow \gamma \psi(2S))=146$ keV and ${\Gamma}({\rm
X(3872)}\rightarrow \gamma \psi(3770))=7.09$ keV. The ratio ${{\rm
Br(X(3872)}\rightarrow\gamma\psi(2{\rm S}))}/{{\rm
Br(X(3872)}\rightarrow \gamma {\sl J}/\psi)}=4.4$ agrees with
experimental data by BaBar, but is larger than the new up-bound
reported by Belle recently. With the same method, we also predicted
the decay widths, ${\Gamma}(\chi_{b1}(1\rm P))\rightarrow \gamma
\Upsilon(1\rm S))=30.0$ keV, ${\Gamma}(\chi_{b1}(2\rm P))\rightarrow
\gamma \Upsilon(1\rm S))=5.65$ keV and ${\Gamma}(\chi_{b1}(2\rm
P))\rightarrow \gamma \Upsilon(2S))=15.8$ keV, from which we get the
full widths: ${\Gamma}(\chi_{b1}(1\rm P))\sim 85.7$ keV and
${\Gamma}(\chi_{b1}(2\rm P))\sim 66.5$ keV.

\end{abstract}
 \maketitle

\section{Introduction}

X(3872) was discovered by Belle Collaboration \cite{Choi} in 2003
through the channel ${\rm B}^{\pm}\rightarrow {\rm K}^\pm\sl
J/\psi\pi^+\pi^-$. The mass reported by Belle is $M=3872.0\pm
0.6\pm0.5$ MeV, and the full width has an upper limit $\Gamma < 2.3$
MeV at 90\% C.L.. Later the existence of this particle was confirmed
by CDF \cite{Acosta}, D0 \cite{Abazov}, and BaBar \cite{Aubert3}
Collaborations.

The radiative decay channel ${\rm X(3872)}\rightarrow \sl
J/\psi\gamma$ \cite{Abe1} indicates that this particle has
positive C-parity. The most possible $\sl J^{PC}$ of X(3872) is
$1^{++}$, which is favored by the analysis of the decay angular
distribution \cite{Abe}. But the dipion mass distribution and
tripion mass distribution in ${\rm X(3872)}\rightarrow \sl
J/\psi\pi^+\pi^-$ and ${\rm X(3872)}\rightarrow \sl
J/\psi\pi^+\pi^-\pi^0$ \cite{Abe1} favor a $\rho$ resonance and a
$\omega$ resonance, respectively, which indicates a large isospin
breaking. The mass of X(3872) is 50-100 MeV smaller than the
predictions of potential models. To understand these puzzles, many
assignments of X(3872) were proposed, besides the traditional
charmonium state assignment \cite{Chao,Barnes,Fazio,Suzuki}, there
are the assignments of a molecular state
\cite{Swanson1,Molecular,zhu,dong2,Valcarce,Braaten}, a hybrid
charmonium \cite{Li}, a diquark-antidiquark state \cite{Maiani},
cusp effect or virtual state \cite{Bugg, Hanhart}(for a review,
see e.g. Ref.~\cite{Swanson}).

If $\rm X(3872)$ is a $1^{++}$ state, then there are two most
possible natural assignments, a molecular state or a traditional
charmonium $\chi_{c1}(2\rm P)$. Molecular state model predicts the
value of $\rm Br(B^+\rightarrow X(3872)K^+)/Br(B^0\rightarrow
X(3872)K^0)$ is about 10\% \cite{Braaten1}, while the experimental
value is $0.5\pm0.3\pm 0.05$ \cite{Aubert4}. This model meets more
serious problems when used to calculate radiative decays. From
Ref.~\cite{Swanson1} (see Table.1), one can get the ratio
$\Gamma_{\psi^\prime\gamma}/\Gamma_{\sl J/\psi\gamma}\sim 4\times
10^{-3}$, while the experimental value by BaBar is $3.4\pm1.4$
\cite{Aubert1}. In this Letter, we will not consider the
possibility of $\rm X(3872)$ as a molecular state, but due to its
E1 radiative decay, we consider the possibility of an ordinary
charmonium state.

Because E1 radiative decays will play a fundamental role in
determination of the nature of $\rm X(3872)$, in this Letter, we
just calculate the radiative E1 decay widths of X(3872) by
assigning it as the $\chi_{c1}$(2P) state and give the results.
Although there is a discrepancy in the mass values of experiments
and models, as Ref.~\cite{Barnes} proposed, this is due to
additional effects, such as coupled-channel effect. For the large
isospin breaking, charmonium model can also give a good
explanation \cite{Chao1}.

This Letter is organized as follows. In Sec.~II, we solve the
instantaneous Bathe-Salpeter (BS) equation (Salpeter equation)
\cite{Salpeter1,Salpeter2}, and get wave functions of initial and
final states. Then within Mandelstam formalism \cite{Mandelstam},
we calculate the transition matrix element. In Sec.~III, we
compare our results with other theoretical predictions and
experimental data, some predictions and discussions are also given
in this section.

\section{E1 decay of X(3872) with $\chi_{c1}$(2P) charmonium assignment}

The wave function of $1^{++}$ sate is ,
\begin{equation}\label{eq1}
\varphi_{1^{+}}(q_{\perp})=i\varepsilon_{\mu\nu\alpha\beta}
P^{\nu}q_{\perp}^{\alpha}\epsilon^{\beta}_1[\varphi_1M\gamma^{\mu}+
\varphi_2{\not\!P}\gamma^{\mu}+\varphi_3{\not\!q}_{\perp}\gamma^{\mu}
+\varphi_4{\not\!P}\gamma^{\mu}{\not\!q}_{\perp}/M]/M^2,
\end{equation}
where $\epsilon_{\mu\nu\alpha\beta}$ is the totally antisymmetric
tensor. $\epsilon_1$ is the polarization vector of the meson while
$M$ is its mass. $\sl P$ and $q$ are the total momentum and relative
momentum of constitute quark and antiquark, respectively, which are
defined as:
\begin{equation}
p_1=\alpha_1 P+q,~\alpha_1=\frac{m_1}{m_1+m_2}, ~p_2=\alpha_2 P-q,~
\alpha_2=\frac{m_2}{m_1+m_2},
\end{equation}
where $p_1, p_2$ are the momenta of quark and antiquark,
respectively. $m_1=m_2$ is the mass of constitute quarks.
$\varphi_i$s are functions of $q_{\perp}^2$. $q_{\perp}$ has the
form: $q_{\perp}^\mu=q^\mu-(P\cdot q/M^2)P^\mu$. Because there are
two constrain conditions \cite{wang}, $\varphi_3$, $\varphi_4$ can
be expressed by $\varphi_1$, $\varphi_2$ \cite{wang}. The wave
function above has a different form with that in \cite{wang}, but
they are equivalent to each other. We show a general wave function
form for $1^+$ state, which means quark and antiquark inside the
meson can have different masses. If we consider charmonium
$1^{++}$ state, the quark and antiquark have the same mass, then
$\varphi_3$ will disappear \cite{wang}.

The wave function of $1^{--}$ state is \cite{wang1},
$$\varphi_{1^{-}}(q'_{\perp})=
q'_{\perp}\cdot{\epsilon}_{2} \left[f_1(q'_{\perp})+{\not\!
P_f}f_2(q'_{\perp})/M_f+
{{\not\!q'_\perp}}f_3(q'_{\perp})/M_f+{{\not\!P_f} {\not\!q'_\perp}}
f_4(q'_{\perp})/{M_f^2}\right]+
M_f{\not\!\epsilon}_{2}f_5(q'_{\perp})$$
\begin{equation}\label{eq2}
+{\not\!\epsilon}_{2}{\not\!P_f}f_6(q'_{\perp})+
({\not\!q'_\perp}{\not\!\epsilon}_{2}-
q'_{\perp}\cdot{\epsilon}_{2})
f_7(q'_{\perp})+({\not\!P_f}{\not\!\epsilon}_{2}
{\not\!q'_\perp}-{\not\!P_f}q'_{\perp}\cdot{\epsilon}_{2})
f_8(q'_{\perp})/M_f,
\end{equation}
where $M_f$, $\sl P_f$ and $\epsilon_2$ are the mass, momentum and
polarization vector of the meson, respectively. Again, if we
consider charmonium, the constitute quark and antiquark inside the
mason have the equal mass. Because there are four constrain
equations~\cite{wang1}, $f_7$ and $f_2$ will disappear, and $f_1$
and $f_8$ can be expressed by $f_3$, $f_4$, $f_5$ and $f_6$. Here we
will not present the details of solving BS equation, which can be
found in Ref.~\cite{wang2}.

We just give the Cornell potential which is applied when solving BS
equation:
\begin{equation}
V(\stackrel\rightarrow{q})=V_s(\stackrel\rightarrow{q})
+\gamma_0\otimes\gamma^0V_v(\stackrel\rightarrow{q}),
\end{equation}
\begin{equation}
V_{s}(\stackrel\rightarrow{q})
=-(\frac{\lambda}{\alpha}+V_0)\delta^{3}(\stackrel\rightarrow{q})
+\frac{\lambda}{\pi^{2}}\frac{1}{(\stackrel\rightarrow{q}^{2}+\alpha^{2})^{2}},
\end{equation}
\begin{equation}
V_v(\stackrel\rightarrow{q})=-\frac{2}{3\pi^{2}}
\frac{\alpha_{s}(\stackrel\rightarrow{q})}{\stackrel\rightarrow{q}^{2}+\alpha^{2}},
\end{equation}
\begin{equation}
\alpha_s(\stackrel\rightarrow{q})=\frac{12\pi}{25}
\frac{1}{{\rm{ln}}(a+\frac{\stackrel\rightarrow{q}^2}{\Lambda_{QCD}})}.
\end{equation}
Here $\lambda$, $\alpha$, e, $V_0$ and $\Lambda_{QCD}$ are
parameters. By fitting the mass spectra of $1^{++}$, $1^{--}$
masons, we can find the best-fit values of these parameters: $a=e=$
2.7183, $\alpha=$ 0.06 GeV, $\lambda=$ 0.2 GeV, $m_c=$ 1.7553 GeV,
$m_b=$ 5.13 GeV, $\Lambda_{QCD}=$ 0.26 GeV ($c\bar c$), 0.20 GeV
($b\bar b$)~(see \cite{wang}). For $1^{++}$ state, $V_0=$-0.452 GeV
($c\bar c$), -0.521 GeV ($b\bar b$), for $1^{--}$ state, $V_0=$
-0.465 GeV ($c\bar c$), -0.570 GeV ($b\bar b$). Here $\alpha$ is the
effective gluon mass. Since the potential we chose is a
phenomenological one and the gluon mass as a parameter is not
running, the value of $\alpha$ here is lower than the usual chosen
especially when it is running close to the infrared limit.

Wave functions above are constructed based on the quantum number
$J^{P}$ or $J^{PC}$ of mesons. For example, $J^{P}$ of every term
in Eq.~(3) is $1^{-}$ (or $1^{--}$ for equal mass system). One can
see that there is S wave and D wave mixing automatically,
especially for the third state ($\psi(3770)$), which is D wave
dominating, but mixing with a small part of S wave. This can be
seen clearly in spherical polar coordinates \cite{wang3},

\begin{figure}
 \centering
\includegraphics[width=5.5in]{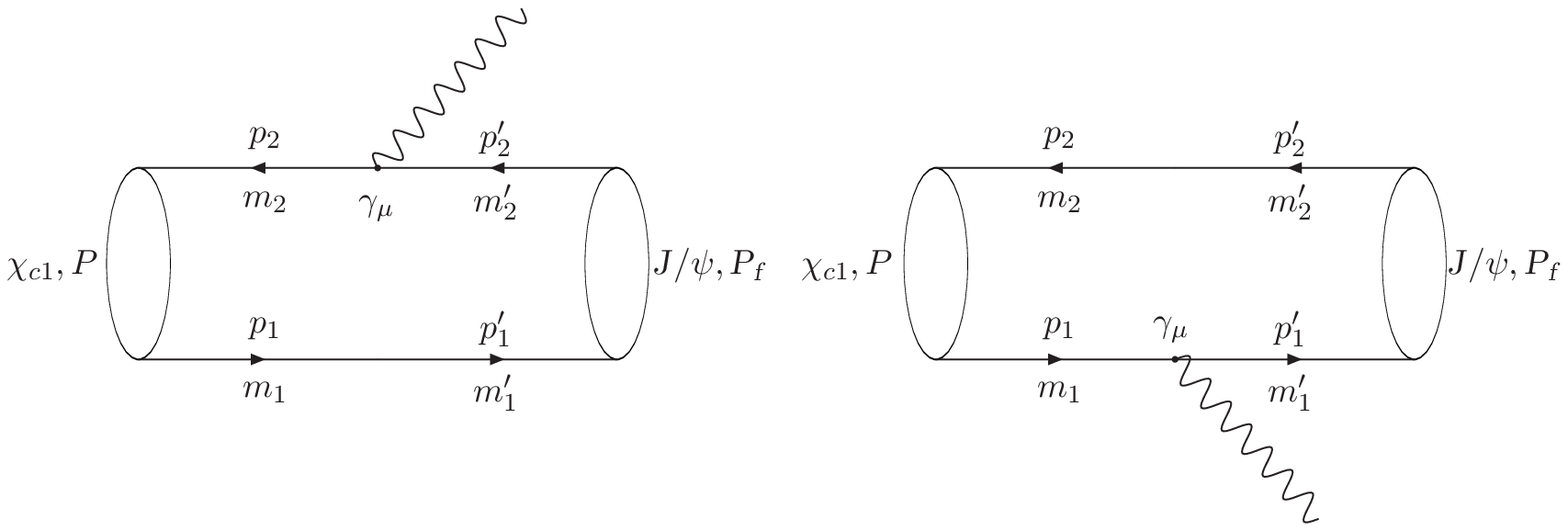}
 \caption{Feynman diagram for the transition: $\chi_{c1}\rightarrow \sl J/\psi + \gamma$. }
\end{figure}

\begin{figure}
 \centering
\includegraphics[width=5.5in]{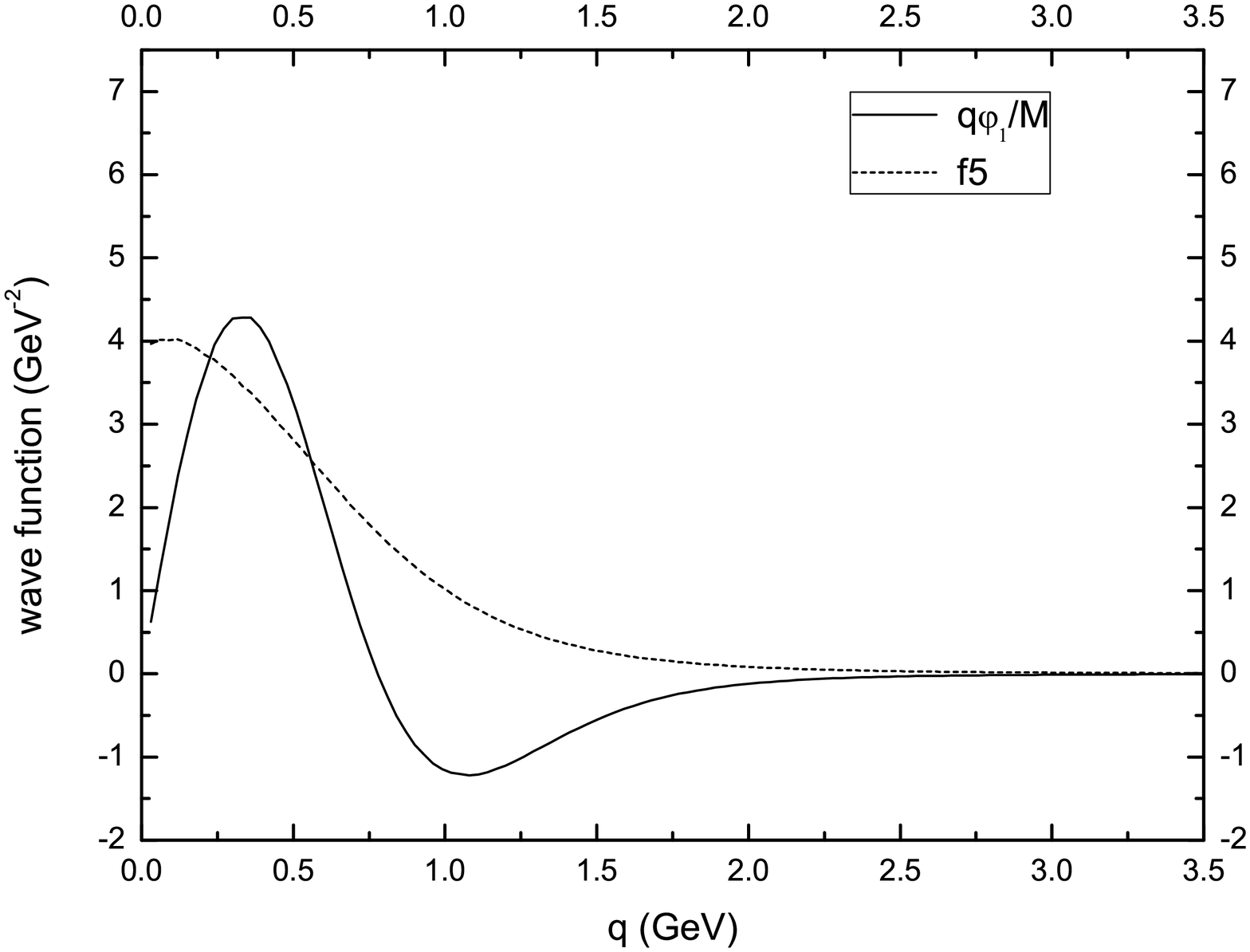}
 \caption{Radial wave function $\frac{|\stackrel{\rightarrow}{q}|}{M}\varphi_1$ of X(3872) and f5 of $\sl J/\psi$.}
\end{figure}

\begin{figure}
 \centering
\includegraphics[width=5.5in]{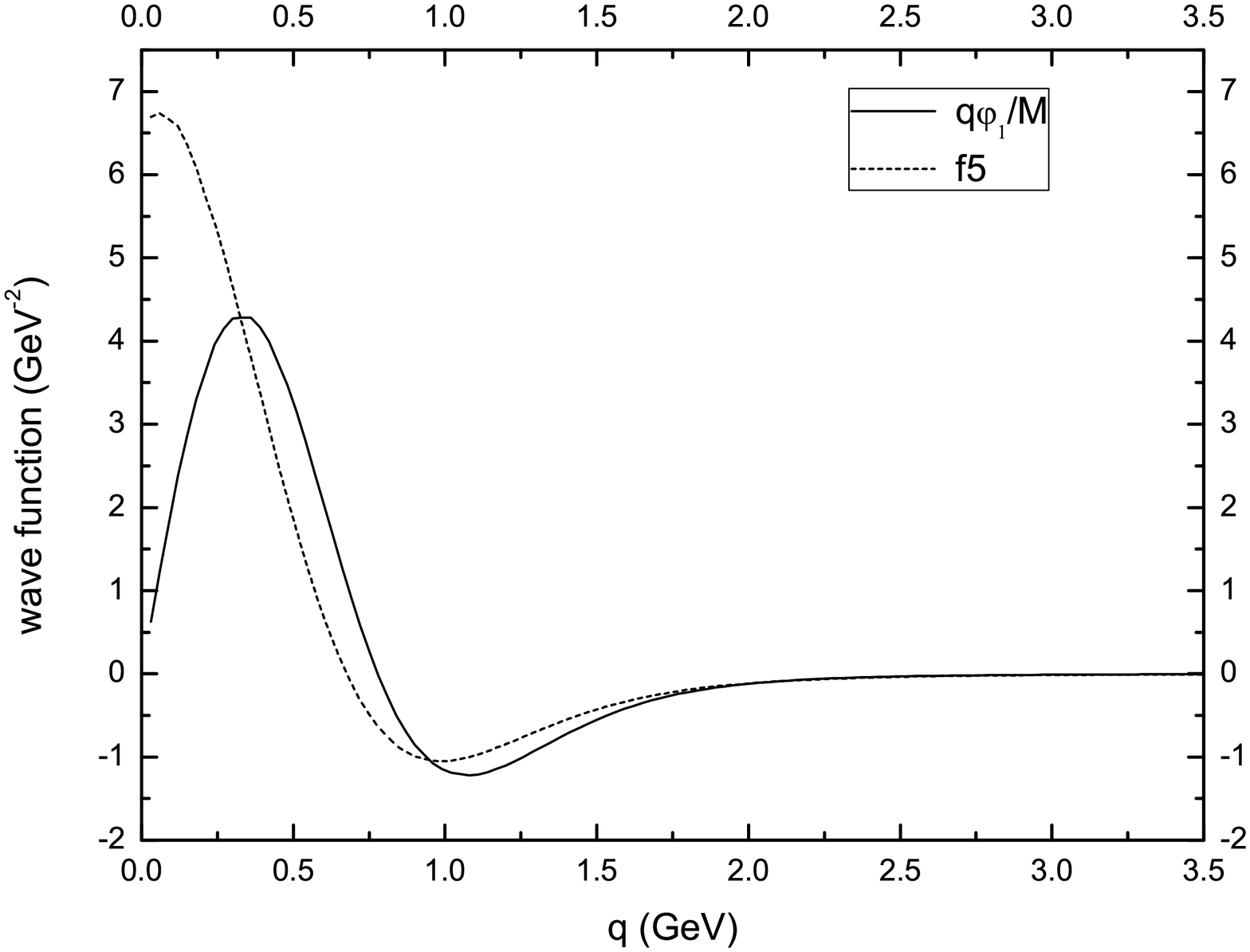}
 \caption{Radial wave function $\frac{|\stackrel{\rightarrow}{q}|}{M}\varphi_1$ of X(3872) and f5 of $\psi(2\rm S)$.}
\end{figure}

\begin{figure}
 \centering
\includegraphics[width=5.5in]{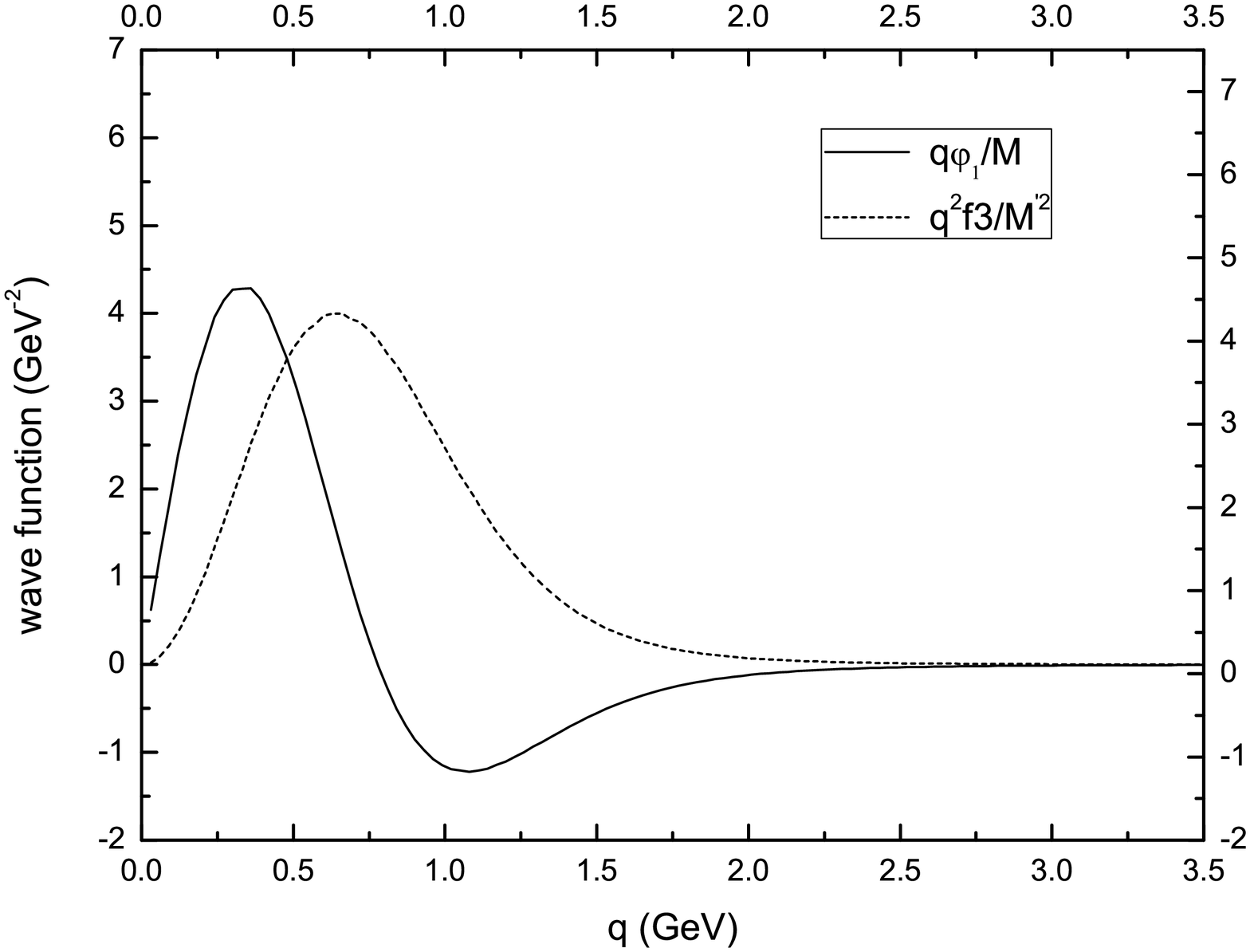}
 \caption{Radial wave function $\frac{|\stackrel{\rightarrow}{q}|}{M}\varphi_1$ of X(3872) and
$\frac{|\stackrel{\rightarrow}{q}|^2}{M^{\prime 2}}$f3 of
$\psi(3770)$.}
\end{figure}

The relativistic transition amplitude of $1^{++}$ state decaying to
a photon and a $1^{--}$ state (see Fig.1) can be written in terms of
BS wave function:
\begin{equation}\label{eq3}
T=\langle P_f \epsilon_2 ,k \epsilon|S|P\epsilon_1\rangle
=\frac{(2\pi)^4ee_q}{\sqrt{2^3\omega_\gamma E
E_f}}\delta^4(P_f+k-P)\epsilon^\xi M_\xi,
\end{equation}
where $\epsilon$, $\epsilon_1$ and $\epsilon_2$ are the polarization
vectors of the photon, initial meson and final meson, respectively.
$P$, $P_f$ and $k$ are the momenta of initial meson, final meson and
photon, respectively. $e_q=\frac{2}{3}$ for charm quark and $e_q=
-\frac{1}{3}$ for bottom quark are the charges in unit of $e$.
$M^\xi$ is the matrix element of the electromagnetic current, which
according to Refs.~\cite{Mandelstam,wang3}, in the leading order
(the order of $\alpha=\frac{e^2}{4\pi}$, also neglect terms contain
$\psi^{+-}$, $\psi^{-+}$ and $\psi^{--}$, which contribute less than
1\%) can be written as:
\begin{equation}\label{eq4}
M^\xi=ee_q\int\frac{d{\stackrel{\rightarrow}{q}}}{(2\pi)^3}
Tr[\frac{{\not\!P}}{M}\bar\varphi^{\prime
++}(q_\bot+\alpha_2P_{f\bot})\gamma^\xi\varphi^{++}(q_{\bot})-\bar\varphi^{\prime
++}(q_\bot-\alpha_1P_{f\bot})\frac{{\not\!P}}{M}\varphi^{++}(q_{\bot})\gamma^\xi],
\end{equation}
where $\varphi^{++}$ is the positive part of BS equation.
$P_{f\perp}$ and $\bar\varphi^{++}$ are defined as
$P_{f\perp}^{\mu}=P_f^{\mu}-(P\cdot P_f/M^2)P^{\mu}$ and $\gamma_0
(\varphi^{++})^{\dagger} \gamma_0$, respectively.

For X(3872), the positive energy part of wave function has the form:
\begin{equation}\label{eq5}
\varphi^{++}_{1^{++}}=i\varepsilon_{\mu\nu\alpha\beta}
P^{\nu}q_{\perp}^{\alpha}\epsilon_{1}^{\beta}(A_{1}\gamma^{\mu}
+A_{2}\gamma^{0}\gamma^{\mu}+A_{3}\gamma^{0}\gamma^{\mu}{\not\!q}_{\perp}),
\end{equation}
where $A_1$, $A_2$, $A_3$ are defined as:
\begin{equation}\label{eq61}
A_1=\frac{1}{2}(\frac{\varphi_1}{M}+\frac{\omega}{m}\frac{\varphi_2}{M}),
\end{equation}
\begin{equation}\label{eq62}
A_2=\frac{1}{2}(\frac{\varphi_1}{M}+\frac{\omega}{m}\frac{\varphi_2}{M})\frac{m}{\omega},
\end{equation}
\begin{equation}\label{eq63}
A_3=\frac{1}{2}(\frac{\varphi_1}{M}+\frac{\omega}{m}\frac{\varphi_2}{M})\frac{1}{\omega}.
\end{equation}

The positive energy part of wave function for $1^{--}$ state can be
written as:
\begin{equation}\label{eq7}
\varphi^{++}_{1^{--}}=B_1 \not\!\epsilon_{2}
+B_2\not\!\epsilon_{2} \not\!P_f +B_3 \not\!P_f\not\!\epsilon_{2}
\not\! q_{\perp}^{\prime}+B_4 q^{\prime}_\perp\cdot \epsilon_{2}
+B_5 q^{\prime}_\perp\cdot \epsilon_{2} \not\!P_f+B_6
q^{\prime}_\perp\cdot \epsilon_{2} \not\!q^\prime_\perp+B_7
q^{\prime}_\perp\cdot \epsilon_{2} \not\!P_f\not\!q^\prime_\perp,
\end{equation}
where the expressions of $B_1$ to $B_7$ can be found in
Ref.~\cite{wang4}.

\begin{table*}[hbt]
\setlength{\tabcolsep}{0.5cm} \caption{\small E1 decay widths of
$\chi_{c1}(1\rm P)$ and $\chi_{c1}(2\rm P)$. In
Ref.~\cite{Barnes}, Barnes and Godfrey (labeled by B\&G) have made
the impulse, nonrelativistic, zero recoil, and dipole
approximations. In Ref.~\cite{Swanson1}, three method are adopted:
one has the same approximations as B\&G, but compute with a
improved potential (labeled by Swanson1); one has no approximation
(labeled by Swanson2); the last one is molecular model (labeled by
Swanson3). Our values inside the parentheses are for the cases
that the mass of $3923$ MeV for $\rm X(3872)$ is chosen.}
\begin{tabular*}{\textwidth}{@{}c@{\extracolsep{\fill}}ccccc}
\hline \hline Ref.&$\Gamma_{\sl J/\psi\gamma}^{\chi_{c1}(1\rm P)}$
(keV) &$\Gamma_{\sl J/\psi\gamma}^{\chi_{c1}(2\rm P)}$ (keV) &$
\Gamma_{\psi(2\rm S)\gamma}^{\chi_{c1}(2\rm P)}$ (keV)&$\Gamma_{\psi(3770)\gamma}^{\chi_{c1}(2\rm P)}$ (keV)\\
\hline {\phantom{\Large{l}}}\raisebox{+.2cm}{\phantom{\Large{j}}}
This work &306&33.0 (33.3)&146 (182)&7.09 (9.83)
\\
{\phantom{\Large{l}}}\raisebox{+.2cm}{\phantom{\Large{j}}}
PDG~\cite{PDG}&320&&&
\\ {\phantom{\Large{l}}}\raisebox{+.2cm}{\phantom{\Large{j}}}
Li and Chao~\cite{Chao} &&45&60&
  \\
  {\phantom{\Large{l}}}\raisebox{+.2cm}{\phantom{\Large{j}}}
 Swanson1~\cite{Swanson1} &&71&95&6.5\\
   {\phantom{\Large{l}}}\raisebox{+.2cm}{\phantom{\Large{j}}}
 Swanson2~\cite{Swanson1} &&139&94&6.4\\
   {\phantom{\Large{l}}}\raisebox{+.2cm}{\phantom{\Large{j}}}
 Swanson3~\cite{Swanson1} &&8&0.03&0\\
{\phantom{\Large{l}}}\raisebox{+.2cm}{\phantom{\Large{j}}}
 B\&G~\cite{Barnes}
&&11.0&63.9&3.7\\
{\phantom{\Large{l}}}\raisebox{+.2cm}{\phantom{\Large{j}}}
Eitchen $\it et$ $\it al$~\cite{Eitchen}&&110&180&25\\

{\phantom{\Large{l}}}\raisebox{+.2cm}{\phantom{\Large{j}}}
Dong $\it et$ $\it al$~\cite{Dong}&&$1\sim2$&$5\sim6$&\\

\hline\hline
\end{tabular*}
\end{table*}

\begin{table*}[hbt]
\setlength{\tabcolsep}{0.5cm} \caption{\small E1 decay widths of
$\chi_{b1}(\rm 1P)$ and $\chi_{b1}(\rm 2P)$.}
\begin{tabular*}{\textwidth}{@{}c@{\extracolsep{\fill}}ccccc}
 \hline \hline Ref.&$\Gamma_{\Upsilon(1\rm S)\gamma}^{\chi_{b1}(1\rm P)}$ (keV)&$\Gamma_{\Upsilon(1\rm S)\gamma}^{\chi_{b1}(2\rm P)}$
 (keV)
&$\Gamma_{\Upsilon(2\rm S)\gamma}^{\chi_{b1}(2\rm P)}$ (keV)\\
\hline {\phantom{\Large{l}}}\raisebox{+.2cm}{\phantom{\Large{j}}}
This work&30.0&5.65&15.8\\
{\phantom{\Large{l}}}\raisebox{+.2cm}{\phantom{\Large{j}}}
Kwong and Rosner~\cite{Kwong}&32.8&9.31&15.9\\
{\phantom{\Large{l}}}\raisebox{+.2cm}{\phantom{\Large{j}}}
Ebert $\it et$ $\it al$~\cite{Ebert}&36.6&7.49&14.7\\
{\phantom{\Large{l}}}\raisebox{+.2cm}{\phantom{\Large{j}}}
Fazio~\cite{Fazio}&107&&\\
  \hline\hline
\end{tabular*}
\end{table*}

\section{Numerical results and discussions}

We first calculated the decay width of $\chi_{c1}(\rm 1P)$ to $\sl
J/\psi$ and $\gamma$. The result $306$ keV shown in Table.1 agrees
with the experimental value $320$ keV very well. This shows that
our method can be used to describe radiative decay. For $\rm
X(3872)$, with the $2{^3\rm P_1}$ charmonium assumption, we
calculated decay widths of three channels. We first solved the
instantaneous BS equation by setting the parameter $V_0=-0.452$
GeV. The mass of $\chi_{c1}$(2P) is $3.923$ GeV \cite{wang}, which
is about $50$ MeV larger than that of X(3872). This is the common
character of all potential models, which may be due to the
coupled-channel effect. The results which we got by using this
wave function are included in parentheses in Table 1. To make the
mass of $\chi_{c1}$(2P) equal to $3872$ MeV, we solved the BS
equation by setting ${\sl V}_0=-0.516$ GeV, and keeping other
parameters un-changed. (This also causes a mass decrease of 50 GeV
for other states. Here we just want to get the wave function of
$\chi_{c1}$(2P) when its mass is $3872$ MeV. To make the spectrum
agree with experimental data, we have to modify our coupled
equations, especially the potential, which is our future work.)
Decay widths with this set of parameters are those outside
parentheses. We can see that the value of $\Gamma_{\sl
J/\psi\gamma}^{\chi_{c1}(2\rm P)}$ is almost unchanged, while the
values of $ \Gamma_{\psi(2\rm S)\gamma}^{\chi_{c1}(2\rm P)}$ and
$\Gamma_{\psi(3770)\gamma}^{\chi_{c1}(2\rm P)}$ are reduced by
nearly 20\% and 30\%, respectively.

One can see that our $\Gamma_{J/\psi\gamma}=33.0$ keV is of the
same order with that of Li and Chao~\cite{Chao},
B\&G~\cite{Barnes} and Swanson1~\cite{Swanson}, but much larger
than $8$ keV of Swanson3~\cite{Swanson1} (molecular model) and
$1\sim2$ keV of Dong~\cite{Dong} (molecular and $\sl c\bar c$
mixture). Our $\Gamma_{\psi^\prime\gamma}=146$ keV is about 2.5
times larger than that of Li and Chao~\cite{Chao} and
B\&G~\cite{Barnes}, but approximately equals to that of
Eitchen\cite{Eitchen}, which has considered the influence of
open-charm channels. The results in Swanson2~\cite{Swanson1} have
used a improved potential and included no zero recoil and dipole
approximation which used in Swanson3~\cite{Swanson1} and
B\&G~\cite{Barnes}. But as B\&G~\cite{Barnes} did, the wave
function and meson mass are calculated by adding spin-dependent
interaction in the Hamiltonian. In this Letter, we started from BS
equation, which is relativistic covariance. By using instantaneous
approximation, we get coupled Salpeter equations, which has
included the relativistic effects automatically.

The ratios of E1 decay widths and the width of ${\rm
X}(3872)\rightarrow \pi^+ \pi^- {\sl J}/\psi$ detected by BaBar are
\cite{Aubert1}:
\begin{equation}\label{eq8}
\frac{{\rm Br}({\rm X}(3872)\rightarrow \gamma \sl J/\psi)}{{\rm
Br}({\rm X}(3872)\rightarrow \pi^{+} \pi^- \sl J/\psi)}=0.33\pm
0.12,
\end{equation}
\begin{equation}\label{eq9}
\frac{\rm Br(\rm X(3872)\rightarrow \gamma \psi(2\rm S))}{{\rm
Br}({\rm X}(3872)\rightarrow \pi^{+} \pi^- \sl J/\psi)}=1.1\pm
0.4.
\end{equation}
Up to now, the widths and branch ratios of E1 decay channels have
not been measured precisely. But the ratio can be drawn from
Eqs.(\ref{eq8}) and Eq.(\ref{eq9}) \cite{Aubert1}:
\begin{equation}\label{eq10}
\frac{{\rm Br(X(3872)}\rightarrow \gamma\psi(2{\rm S}))}{{\rm
Br(X(3872)}\rightarrow \gamma \sl J/\psi)}=3.4\pm 1.4.
\end{equation}
With our results in Table 1 we get this ratio:
\begin{equation}\label{eq11}
\frac{{\rm Br(X(3872)}\rightarrow\gamma \sl \psi(2{\rm S}))}{{\rm
Br(X(3872)}\rightarrow \gamma {\sl J}/\psi)}=4.4,
\end{equation}
which is very close to that of Eq.~(\ref{eq10}). In
Refs.~\cite{Chao} and \cite{Barnes} this ratio is 1.3 and 6.1,
respectively. We can see that models with charmonium assumption
can predict this ratio correctly, while molecular model prediction
is very small. In Ref.~\cite{Dong}, a composite state which
contains both molecular hadronic component and a $\sl c\bar c$
component was considered. By changing the mixing angle, a correct
ratio can be reached, but the decay widths are dramatically
changed.

Recently Bhardwaj reported the new results of Belle on ${\rm
X}(3872)$ at the QWG2010 conference \cite{vishal}, which is
$\label{eq11} \frac{{\rm Br(X(3872)}\rightarrow\gamma \sl \psi(2{\rm
S}))}{{\rm Br(X(3872)}\rightarrow \gamma {\sl J}/\psi)}<2.1 $. Our
result with the $\chi_{c1}(2\rm P)$ assignment is two times larger
than this up-bound, so there is still long way to go to know the
nature of ${\rm X}(3872)$.

The large ratio $\rm
\Gamma_{\psi^\prime\gamma}/\Gamma_{J/\psi\gamma}$ can be
understood by Figs. 2 and 3. For $\sl J/\psi$, its wave function
has no node, that is the numerical values of the wave function in
the whole space are all positive (see Fig.2), while for $\rm
X(3872)$ and $\psi(2S)$, since they are the radial excited states
of $\chi_{c1}(\rm 1P)$ and $J/\psi$, respectively, the wave
functions have one node, that is, before the node the values of
wave functions are positive, after the node the values are
negative. So when we calculate the transition amplitude, we need
to compute the overlap integral shown in Eq.~(\ref{eq4}). There
exists dramatically cancellation in the overlap integral before
the node and after the node when we consider the decay ${\rm
X(3872)}\rightarrow \gamma {\sl J}/\psi$ which can be seen from
Fig.2. This is the reason why the decay width $33.0$ keV of this
channel is much smaller (almost one order) than the width $306$
keV of channel $\chi_{c1}({\rm 1P})\rightarrow \gamma {\sl
J}/\psi$. But for the decay $\rm X(3872)\rightarrow \gamma
\psi(2\rm S)$, the two overlapping wave functions both have the
node structures, see Fig.3. So only in the region where one wave
function is before the node, while the other is after the node,
the overlapping integral gives negative contributions. And we can
see from Fig. 3 that only a very small part of phase space will
give negative contributions, so there is almost no cancellation
when we calculate the transition amplitude. Finally we get a large
decay width $146$ keV for the channel of $\rm X(3872)\rightarrow
\gamma \psi(2\rm S)$.

We have mentioned that the numerical values of $ \Gamma_{\psi(2\rm
S)\gamma}^{\chi_{c1}(2\rm P)}$ and
$\Gamma_{\psi(3770)\gamma}^{\chi_{c1}(2\rm P)}$ are very sensitive
to the mass of $\rm X(3872)$ (see Table 1). This can be explained
by different phase space and the node structure of wave functions.
From Eq.~(\ref{eq4}) we can see that in the overlap integral the
relative momentum $\stackrel\rightarrow q_\perp$ of final state
has a shift $\alpha_2\stackrel\rightarrow{P_f}$ or
$-\alpha_1\stackrel\rightarrow{P_f}$. When we change the mass of
$\chi_{c1}(2\rm P)$ from $3872$ to $3923$ MeV, the node position
in the wave functions has almost no change, but the value of
$|\stackrel\rightarrow{P_f}|$ will change obviously due to
different phase space, for example, from $181$ MeV to $230$ MeV
for $ {\chi_{c1}(2\rm P)}\rightarrow{\psi(2\rm S)\gamma}$, which
also changes the overlap integral. Finally we got much different
values of decay width $ \Gamma_{\psi(2\rm
S)\gamma}^{\chi_{c1}(2\rm P)}$. Similar conclusion can be obtained
for the case of $\chi_{c1}(2\rm P)\rightarrow \gamma \psi(3770)$
(see Fig.4). But for ${\chi_{c1}(2\rm P)}\rightarrow \gamma {\sl
J}/\psi$, the relative small mass of $J/\psi$ results in similar
values of $|\stackrel\rightarrow{P_f}|$ for both cases, $695$ MeV
and $736$ MeV, so the decay widths are similar for both cases. The
two-body decay width can be written as: $\Gamma=\frac{1}{8\pi
M}\frac{|\stackrel\rightarrow P_f|}{M}\bar\Sigma|T|^2$. So the
pure change caused by the change of phase space is 23.8\% for $
{\chi_{c1}(2\rm P)}\rightarrow{\psi(2\rm S)\gamma}$ and 3.2\% for
${\chi_{c1}(2\rm P)}\rightarrow \gamma {\sl J}/\psi$. From Table 1
the total change of the two processes is 24.7\% and 0.9\%
respectively, which means most of the change for $ {\chi_{c1}(2\rm
P)}\rightarrow{\psi(2\rm S)\gamma}$ comes from phase space while
for ${\chi_{c1}(2\rm P)}\rightarrow \gamma {\sl J}/\psi$ the
larger contribution comes from the change of matrix element.

Using the same method, we also calculated the radiative E1 decay
widths of $\chi_{b1}(1\rm P)$ and $\chi_{b1}(2\rm P)$, and we show
the results predicted by our method and other models in Table.2. One
can see that the decay width $\Gamma(\chi_{b1}(1\rm P)\rightarrow
\gamma\Upsilon (1S))=30.0$ keV calculated by our method is about 3
times smaller than that of Refs.~\cite{Fazio}, but close to the
values in Ref.~\cite{Kwong} and Ref.~\cite{Ebert}, which are $32.8$
keV and $36.6$ keV, respectively. There are still no experimental
data of these radiative decay widths, however, ratios are available.
Particle Data Group \cite{PDG} has listed the branching ratios: $\rm
Br(\chi_{b1}(1\rm P)\rightarrow \gamma\Upsilon (1\rm S))=(35\pm
8)\times 10^{-2}$, $\rm Br(\chi_{b1}(2\rm P)\rightarrow
\gamma\Upsilon (1\rm S))=(8.5\pm 1.3)\times 10^{-2}$, $\rm
Br(\chi_{b1}(2\rm P)\rightarrow \gamma\Upsilon (2\rm S))=(21\pm
4)\times 10^{-2}$, so from this experimental data, we can get the
ratio \cite{Fazio}:
\begin{equation}\label{eq12}
\frac{{\rm Br}(\chi_{b1}(2\rm P)\rightarrow \gamma \Upsilon (2\rm
S))}{{\rm Br}(\chi_{b1}(2\rm P)\rightarrow \gamma \Upsilon (1\rm
S))}=2.5\pm 0.6.
\end{equation}
Our result is
\begin{equation}\label{eq13}
\frac{{\rm Br}(\chi_{b1}(2\rm P)\rightarrow \gamma \Upsilon (2\rm
S))}{{\rm Br}(\chi_{b1}(2\rm P)\rightarrow \gamma \Upsilon (1\rm
S))}=2.8.
\end{equation}
One can see that it's agreeable with the experimental value. The
full decay widths of $\chi_{b1}$(1P) and $\chi_{b1}$(2P) can be
estimated by the branching ratios and our predicted decay widths.
The results are: $\Gamma_{\chi_{b1}(1\rm P)}\sim$ 85.7 keV, and
$\Gamma_{\chi_{b1}(2\rm P)}\sim$ 66.5 keV.

In conclusion, we first calculated the radiative E1 decay width of
$\chi_{c1}$(1P). The excellent agreement between our result and
experimental value shows that this method we used is good to deal
with the charmonium radiative decays. Then with the traditional
radial excited charmonium state $\chi_{c1}$(2P) assignment for
$\rm X(3872)$ we calculated the radiative E1 decay widths of this
particle, ${\Gamma}({\rm X(3872)}\rightarrow \gamma \sl
J/\psi)=33.0$ keV, ${\Gamma}({\rm X(3872)}\rightarrow \gamma
\psi(2S))=146$ keV and ${\Gamma}({\rm X(3872)}\rightarrow \gamma
\psi(3770))=7.09$ keV. The value of
$\Gamma_{\psi^\prime\gamma}/\Gamma_{\psi\gamma}$ is 4.4, which is
consistent with experimental result by BaBar, but is larger than
the up-bound reported by Belle recently.

We also estimated the radiative E1 decay widths of the bottomonia
states $\chi_{b1}(1\rm P)$ and $\chi_{b1}(2\rm P)$. Results are
$\Gamma(\chi_{b1}(1\rm P)\rightarrow\gamma\Upsilon(1\rm S) )=30.0$
keV, $\Gamma(\chi_{b1}(2\rm P)\rightarrow\gamma\Upsilon(1\rm
S))=5.65$ keV, and $\Gamma(\chi_{b1}(2\rm
P)\rightarrow\gamma\Upsilon(2\rm S))=15.8$ keV. The predicted
ratio $\Gamma_{\Upsilon^\prime\gamma}/\Gamma_{\Upsilon\gamma}$ of
$\chi_{b1}(2\rm P)$ is consistent with experimental data. The full
decay widths of $\Gamma(\chi_{b1}(1\rm P))=85.7$ keV and
$\Gamma(\chi_{b1}(2\rm P))=66.5$ keV (by the channel
$\chi_{b1}(2\rm P)\rightarrow\gamma\Upsilon(1\rm S)$) are also
estimated.

\section*{Acknowledgments}
We would like to thank Chang-Zheng Yuan for his helpful discussion
and reminding us the new results by Belle. This work was supported
in part by the National Natural Science Foundation of China (NSFC)
under Grant No. 10875032 and in part by Projects of International
Cooperation and Exchanges NSFC under Grant No. 10911140267.

\end{document}